\documentclass[manuscript,screen]{acmart}
\usepackage{listings}
\usepackage{xcolor}

\definecolor{codegray}{gray}{0.96}

\lstdefinestyle{skywing}{
	language=Python,
	basicstyle=\ttfamily\footnotesize,
	backgroundcolor=\color{codegray},
	frame=single,
	rulecolor=\color{black!25},
	framerule=0.4pt,
	framesep=6pt,
	xleftmargin=0.5em,
	xrightmargin=0.5em,
	aboveskip=8pt,
	belowskip=8pt,
	columns=fullflexible,
	keepspaces=true,
	showstringspaces=false,
	breaklines=true,
	tabsize=4
}

\AtBeginDocument{%
  }

\setcopyright{none}
\begin{document}

%%
%% The "title" command has an optional parameter,
%% allowing the author to define a "short title" to be used in page headers.
\title{Skywing: A Platform for Decentralized Mathematical Computing in Unreliable Environments}

%%
%% The "author" command and its associated commands are used to define
%% the authors and their affiliations.
%% Of note is the shared affiliation of the first two authors, and the
%% "authornote" and "authornotemark" commands
%% used to denote shared contribution to the research.
\author{Alyson Fox}
\email{fox33@llnl.gov}
%\orcid{1234-5678-9012}
\affiliation{%
	\institution{Lawrence Livermore National Laboratory}
	\city{Livermore}
	\state{California}
	\country{United States}
}

\author{Colin Ponce}
%\orcid{1234-5678-9012}
\affiliation{%
  \institution{Lawrence Livermore National Labratory}
  \city{Livermore}
  \state{California}
  \country{United States}
}
\email{ponce11@llnl.gov}

\author{Annika Mauro}
%\orcid{1234-5678-9012}
\affiliation{%
  \institution{Lawrence Livermore National Labratory}
  \city{Livermore}
  \state{California}
  \country{United States}
}
\email{mauro3@llnl.gov}

\author{Wayne Mitchell}
%\orcid{1234-5678-9012}
\affiliation{%
  \institution{Lawrence Livermore National Labratory}
  \city{Livermore}
  \state{California}
  \country{United States}}
\email{mitchell82@llnl.gov}

\author{Sarah Osborn}
%\orcid{1234-5678-9012}
\affiliation{%
	\institution{Lawrence Livermore National Labratory}
	\city{Livermore}
	\state{California}
	\country{United States}}
\email{osborn9@llnl.gov}

\author{Tom Benson}
%\orcid{1234-5678-9012}
\affiliation{%
	\institution{Lawrence Livermore National Labratory}
	\city{Livermore}
	\state{California}
	\country{United States}}
\email{benson31@llnl.gov}

\author{Shayna Kapadia}
%\orcid{1234-5678-9012}
\affiliation{%
\institution{Lawrence Livermore National Laboratory}
	\city{Livermore}
	\state{California}
	\country{United States}}
\email{kapadia2@llnl.gov}

%%
%% By default, the full list of authors will be used in the page
%% headers. Often, this list is too long, and will overlap
%% other information printed in the page headers. This command allows
%% the author to define a more concise list
%% of authors' names for this purpose.
\renewcommand{\shortauthors}{Fox et al.}

%%
%% The abstract is a short summary of the work to be presented in the
%% article.
\begin{abstract}
Emerging edge, autonomous, and cyber-physical systems increasingly require mathematical computation across heterogeneous devices connected by unreliable communication networks. 
Traditional high-performance computing and distributed data-processing frameworks provide powerful abstractions for managed environments but are less suited to decentralized settings where centralized coordination, reliable communication, and global synchronization cannot be assumed.
This paper presents Skywing, an open-source platform for decentralized mathematical computing in unreliable environments. 
Its programming model consists of three abstractions: agents represent participants in a decentralized computation, processors encapsulate algorithm-specific update rules, and iterations manage distributed execution. 
Skywing supports asynchronous operation, publish-subscribe communication, managed message handling, and the composition of independent algorithms into complex decentralized workflows.
We demonstrate Skywing using representative algorithms from consensus, optimization, and numerical linear algebra. 
Experiments on the native Skywing runtime include Push Sum and Max Consensus, a composed monitoring and control workflow, resilient Push Sum under delayed communication, and resilient asynchronous Jacobi under malevolent data corruption. 
These demonstrations show that Skywing supports diverse decentralized algorithms while separating mathematical logic from communication and execution infrastructure. 
Skywing serves as both a deployment framework for decentralized applications and a research platform for developing resilient mathematical algorithms.
\end{abstract}

%%
%% The code below is generated by the tool at http://dl.acm.org/ccs.cfm.
%% Please copy and paste the code instead of the example below.
%%


%%
%% Keywords. The author(s) should pick words that accurately describe
%% the work being presented. Separate the keywords with commas.

%\received{2 September 2026}
%\received[revised]{12 March 2009}
%\received[accepted]{5 June 2009}

%%
%% This command processes the author and affiliation and title
%% information and builds the first part of the formatted document.
\maketitle
%%
%% If your work has an appendix, this is the place to put it.
%\appendix

\section{Introduction}

Emerging computing applications increasingly rely on collections of distributed devices that must sense, communicate, and make decisions outside the tightly controlled environments assumed by traditional high-performance and cloud computing systems.
Examples include distributed energy resource coordination, autonomous sensing and control with drones or robotic agents, sensor networks, and edge computing systems in which computation must occur close to the physical process being monitored or controlled \cite{erlandson2023resilient_sacd,fox2022algorithmic_unreliable,shi2016edge,satyanarayanan2017edge}. 
In these settings, participating devices may differ substantially in computational capability, communicate over unreliable networks, and operate continuously on streaming data rather than as part of a single terminating batch computation. \cite{erlandson2023resilient_sacd,fox2022algorithmic_unreliable,shi2016edge,bonomi2012fog}.

These environments introduce challenges that are difficult to accommodate with conventional distributed computing tools. 
 Communication may be delayed, intermittent, or lossy. 
 Devices may disconnect, fail, or operate at different rates. Data may become stale or corrupted, and some agents may behave unpredictably or maliciously during operation \cite{vogl2024async_jacobi_resilience,erlandson2023resilient_sacd,fox2022algorithmic_unreliable}.
 Even when failures are not catastrophic, asynchronous arrivals, heterogeneous compute rates, and changing network conditions complicate the implementation of mathematical algorithms that were originally designed for reliable, synchronized, and relatively homogeneous systems \cite{frommer2000asynchronous,hookdingle2018asyncjacobi,wolfsonpou2018convergence,wolfsonpou2019modeling}.

This creates a practical gap for both algorithm researchers and application developers. 
Mature high-performance computing frameworks provide powerful abstractions for large-scale numerical computation, but they generally assume coordinated execution within well-managed computing environments \cite{mpi_standard,openmp_standard}.
Distributed data platforms support scalable data processing, but their centralized orchestration and batch-oriented execution models are poorly matched to continuously executing peer-to-peer numerical algorithms \cite{dean2004mapreduce,zaharia2010spark,zaharia2012rdd,white2012hadoop}. 
Meanwhile, many decentralized approaches, including consensus, federated learning, distributed optimization, and multi-agent control, provide important algorithmic foundations but are often implemented as specialized systems or one-off research prototypes \cite{mcmahan2017federated,kairouz2021federated,olfati2007consensus,boyd2006gossip}.
What is missing is a reusable software platform designed specifically for developing, composing, and deploying decentralized mathematical algorithms in unreliable environments \cite{fox2022algorithmic_unreliable}.

This paper presents Skywing, an open-source platform for decentralized mathematical computing in unreliable environments \cite{skywing_software}.
Skywing provides a reusable programming model for implementing distributed algorithms without requiring a central coordinator. 
The platform is organized around three core abstractions. 
An \emph{Agent} represents a participant in a decentralized computation and manages local communication and execution resources. 
Here, \emph{Agent} refers specifically to the Skywing software abstraction and should not be confused with the LLM based agents commonly used in AI systems.
A {\it Processor} encapsulates the mathematical update rule of an algorithm. 
An {\it Iteration} combines an agent and processor into an asynchronously executing distributed task. 
Processors exchange information through publish-subscribe tags, allowing algorithms to communicate through named data channels without requiring explicit communication logic within processor implementations.

\begin{figure*}[h!] 
	\centering
	 \includegraphics[width=.8\textwidth]{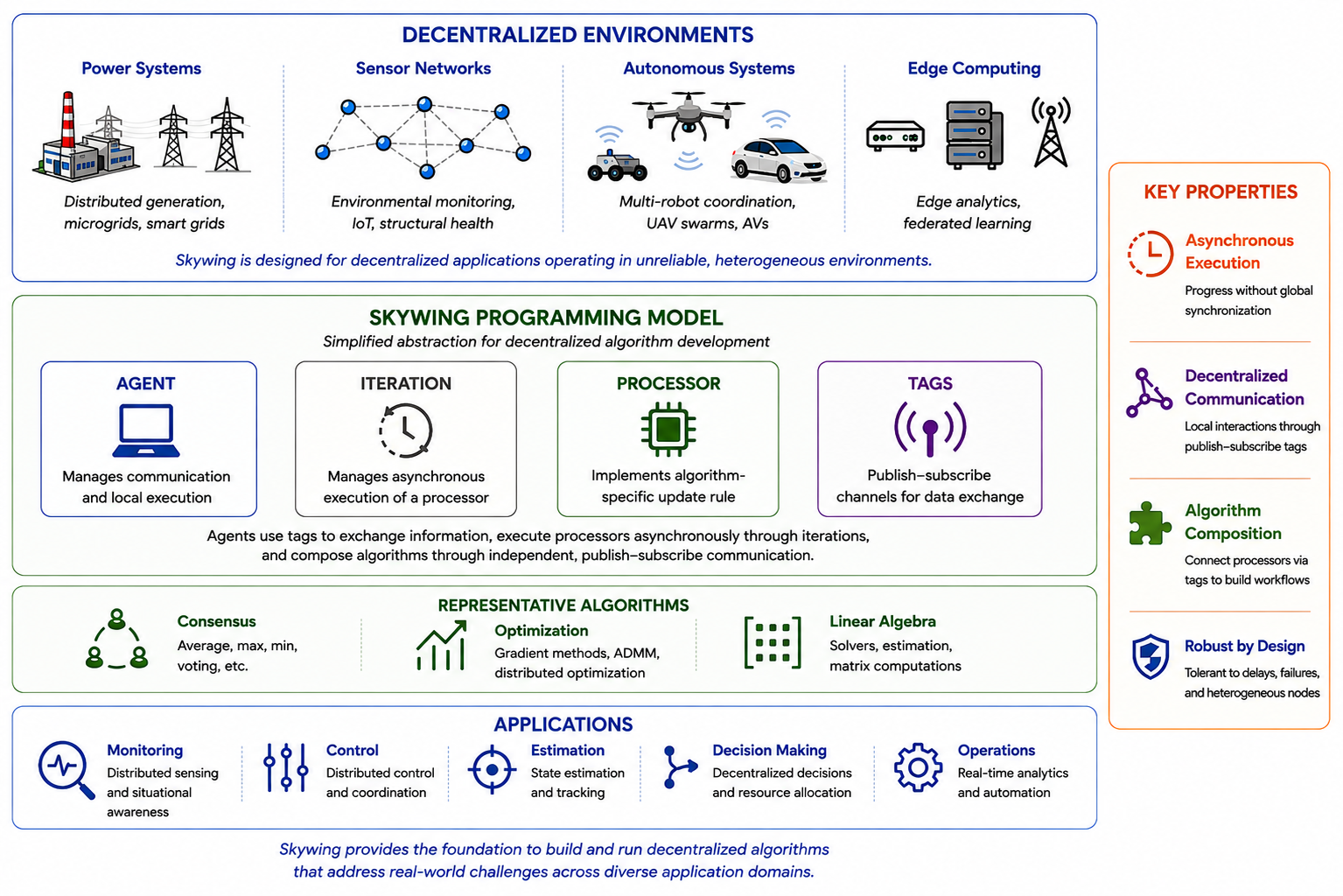} 
	 \caption{Overview of the Skywing platform. Skywing provides a common programming model for developing decentralized mathematical algorithms across diverse operating environments, including power systems, sensor networks, autonomous systems, and edge computing. The platform is built around the Agent--Iteration--Processor abstraction and supports asynchronous execution, decentralized communication, algorithm composition, and deployment in unreliable environments.} 
	 \label{fig:skywing_overview} 
 \end{figure*}
Figure~\ref{fig:skywing_overview} summarizes the role of Skywing within the broader decentralized computing ecosystem. Applications such as distributed energy systems, autonomous systems, sensor networks, and edge computing require mathematical computation across collections of communicating devices.
Skywing provides a common programming and execution environment for these settings, supporting algorithm classes that include consensus, optimization, and numerical linear algebra. 
Figure~\ref{fig:skywing_overview} illustrates how the Skywing programming model connects these algorithmic building blocks to broader decentralized application domains.

The design of Skywing is guided by three goals. 
First, Skywing separates algorithm-specific computation from communication and execution infrastructure, allowing developers to implement mathematical update rules without repeatedly building distributed systems machinery. 
Second, it supports asynchronous execution so that participating agents can make progress without global
synchronization. Third, it supports algorithm composition by allowing multiple processors to execute as independent computational components and be coordinated within larger decentralized workflows.

Our motivating application domain is decentralized monitoring and control for power systems with distributed energy resources, where centralized coordination may be unavailable, undesirable, or vulnerable to failure \cite{duf2016distributed_der,erli2017distributed_opf,zhang2018decentralized_grid,vogl2024async_jacobi_resilience}. 
However, the need for this type of software infrastructure extends beyond the power grid. 
Similar requirements arise in autonomous systems, sensor networks, robotic teams, distributed scientific instruments, and heterogeneous edge computing environments \cite{erlandson2023resilient_sacd,shi2016edge,satyanarayanan2017edge}. 
In each case, the central challenge is not simply to execute an algorithm across many devices, but to support mathematical computation under unreliable communication, asynchronous progress, and changing local conditions.

Skywing has already served as the implementation platform for prior work on asynchronous and resilient iterative numerical algorithms, including asynchronous Jacobi variants and resilient s-step asynchronous conjugate direction methods for solving systems of linear equations \cite{vogl2024async_jacobi_resilience,erlandson2023resilient_sacd}.  
This paper shifts the focus from individual algorithms to the platform itself. 
The following sections describe Skywing’s programming model, software architecture, and execution abstractions, and demonstrate how the same runtime supports consensus algorithms, composed decentralized workflows, resilient communication protocols, and decentralized numerical linear algebra.

The main contribution of this paper is the presentation of Skywing as a general software platform for decentralized mathematical computing. 
Rather than replacing traditional high-performance computing frameworks in the environments for which they are well suited, Skywing targets a different point in the design space: unreliable, heterogeneous, coordinator-free systems where synchronization is expensive, failures are expected, and algorithms must continue operating through local communication.
We describe Skywing's programming model and show how the Agent, Processor, and Iteration abstractions separate algorithm-specific computation from communication and execution infrastructure. 
We demonstrate the breadth of this model through implementations spanning consensus, optimization, and numerical linear algebra, as well as through the composition of independent processors into larger computational workflows. Representative experiments then demonstrate these capabilities under native distributed execution, delayed communication, and adversarial data corruption.
Together, these results position Skywing as both a deployment framework for decentralized applications and a research platform for developing and evaluating mathematical algorithms in unreliable computing environments.
\section{Related Work and Positioning}

\subsection{Distributed computing platforms}

Skywing is related to a broad ecosystem of distributed computing platforms, including high-performance computing frameworks, distributed data processing systems, and edge computing runtimes. 
In scientific computing, MPI and OpenMP are widely used to implement high-performance numerical algorithms \cite{mpi_standard,openmp_standard}. 
These frameworks are highly effective in managed computing environments, especially where resources are relatively homogeneous, communication is reliable, and synchronization can be coordinated. 

A large body of work has also sought to reduce communication and synchronization costs within this model, including communication-avoiding and pipelined Krylov methods \cite{hoemmen2010ca_krylov,ghysels2014pipelinedcg}. 
However, these approaches still largely assume stable execution environments and do not directly provide a software substrate for decentralized, continuously operating computation across unreliable devices.
Skywing complements these frameworks by targeting decentralized mathematical computations in which peer to peer asynchronous execution is a fundamental requirement \cite{shi2016edge,satyanarayanan2017edge,fox2022algorithmic_unreliable}.   
Rather than assuming coordinated progress, its programming model allows agents to operate independently using information available from neighboring agents.

%The operating assumptions targeted by Skywing are different. 
%Skywing is designed for decentralized environments in which no central scheduler or coordinator can be assumed, devices may operate at different speeds, and communication may be delayed, intermittent, or unreliable \cite{shi2016edge,satyanarayanan2017edge,fox2022algorithmic_unreliable}. 
%In such settings, the programming model must allow agents to make progress independently using the most recent information available from their neighbors. Skywing therefore complements traditional HPC frameworks rather than replacing them: it supports a class of distributed mathematical computations where peer-to-peer asynchronous execution is a central requirement rather than an implementation detail.

Distributed data platforms such as MapReduce, Hadoop, and Spark provide another important point of comparison. 
These systems support scalable computation over distributed resources, but they are primarily designed around centralized orchestration, batch analytics, and dataflow-style execution \cite{dean2004mapreduce,white2012hadoop,zaharia2010spark,zaharia2012rdd}. 
Their abstractions are well suited to large-scale data processing, but less natural for decentralized iterative numerical algorithms whose state evolves continuously through local communication among neighboring agents. 
Skywing instead treats each participant as an autonomous computational agent and provides runtime support for continuously executing mathematical processors that exchange information through publish-subscribe channels.

\subsection{Asynchronous, decentralized, and resilient algorithms}

Skywing is directly motivated by work on asynchronous iterative methods, decentralized optimization, consensus algorithms, and resilient numerical computation. Classical work on chaotic relaxation and asynchronous iterations established that certain iterative methods can converge without strict global synchronization \cite{chazan1969chaotic,baudet1978asynchronous,bertsekas1989parallel}. 
Subsequent work on asynchronous Jacobi methods, asynchronous multigrid, and related numerical algorithms further developed the analysis and performance understanding of methods whose updates arrive at different times and whose convergence depends on nonuniform communication patterns \cite{frommer2000asynchronous,hookdingle2018asyncjacobi,wolfsonpou2018convergence,wolfsonpou2019modeling,wolfsonpou2019multigrid}. 
These studies motivate the development of a programming model that can support algorithms whose communication and progress are inherently asynchronous.

Decentralized optimization and consensus methods provide another foundation for Skywing. 
Gossip algorithms, distributed subgradient methods, ADMM-based coordination, and multi-agent consensus methods all study computation over networks without a single centralized controller \cite{olfati2007consensus,boyd2006gossip,nedic2009distributed,boyd2011admm,tsitsiklis1984decentralized}. 
Robust variants of decentralized optimization and ADMM have also been developed for unreliable or adversarial settings \cite{li2018robustadmm}. 
These methods are important algorithmic building blocks for many of the applications Skywing targets, including distributed state estimation, resource allocation, control, and learning. 
However, much of this work focuses on the design and analysis of particular algorithms rather than on reusable software infrastructure for implementing, composing, and deploying broad classes of decentralized mathematical methods.

Resilience is also central to the Skywing design space. Traditional fault tolerance techniques in high-performance and cloud computing often rely on checkpointing, redundancy, restart, or centralized recovery mechanisms \cite{cappello2009fault,huang1984abft}. 
These strategies can be effective in managed environments, but they are less directly applicable to edge and cyber-physical settings where storage may be limited, synchronization is expensive, and devices may join, leave, or fail unpredictably \cite{shi2016edge,satyanarayanan2017edge,fox2022algorithmic_unreliable}. 
Algorithm-based resilience provides a complementary approach by incorporating fault handling into the mathematical method itself rather than relying solely on external recovery mechanisms \cite{huang1984abft,vogl2024async_jacobi_resilience,erlandson2023resilient_sacd}.

Recent work has shown that asynchronous numerical methods can remain effective even in unreliable and heterogeneous computing environments \cite{fox2022algorithmic_unreliable,vogl2024async_jacobi_resilience,erlandson2023resilient_sacd}. 
Prior work from our group includes asynchronous Jacobi methods and resilient variants designed to tolerate corrupted data arising from bit flips, communication errors, or malicious interference \cite{vogl2024async_jacobi_resilience}. 
It also includes s-ACD, an asynchronous collaborative method for solving systems of linear equations, along with resilient extensions that address related data-corruption failures \cite{erlandson2023resilient_sacd}. 
These studies demonstrate the feasibility of decentralized numerical computation without strict synchronization, but they focus on particular algorithms rather than on the reusable software platform needed to support a broader class of methods.

Skywing does not attempt to solve resilience entirely at the runtime level. Instead, it provides a software platform in which resilient algorithms can be implemented and evaluated under common execution conditions. 
The runtime manages communication, asynchronous execution, and process coordination, while resilience mechanisms remain encapsulated within processor implementations. 
This separation allows researchers to compare alternative algorithmic resilience strategies without modifying the underlying distributed infrastructure.

\subsection{Skywing's role as software infrastructure}
Skywing's role as software infrastructure is distinguished by its emphasis on reusable programming abstractions for decentralized mathematical algorithms. 
Unlike domain specific systems in robotics, autonomy, and distributed control, Skywing is intended to support a broad range of mathematical methods rather than a particular control architecture or application domain \cite{olfati2007consensus,ren2007information,oh2015survey}.

The closest prior work from the authors' group used Skywing as an implementation platform for specific resilient asynchronous algorithms, including asynchronous Jacobi variants and s-ACD-based solvers \cite{vogl2024async_jacobi_resilience,erlandson2023resilient_sacd}. 
Those studies demonstrated the feasibility of decentralized numerical computation under unreliable conditions, but they did not present Skywing itself as the primary contribution. 
The contribution of the present paper is to describe the underlying platform: its programming abstractions, execution model, support for publish-subscribe communication, and ability to compose independent algorithms into larger decentralized workflows \cite{skywing_software}.

This positioning is important because many emerging applications require more than a single decentralized algorithm. A power grid monitoring application, for example, may combine distributed aggregation, optimization, local control, and safety logic \cite{duf2016distributed_der,erli2017distributed_opf,zhang2018decentralized_grid}. 
A robotic or sensing application may combine consensus, estimation, planning, and decision-making \cite{olfati2007consensus,ren2007information,oh2015survey}. 
Skywing provides a common execution environment in which these components can be implemented as independent processors and coordinated as part of larger computational workflows.
As a result, decentralized algorithms become reusable computational building blocks rather than standalone distributed programs.

%In this sense, Skywing is both a research platform and a deployment framework. 
%It enables algorithm developers to prototype new decentralized mathematical methods, compare them under common runtime conditions, introduce controlled communication failures or data corruption, and compose multiple methods into larger applications. 
%At the same time, its coordinator-free execution model and communication abstractions are designed for real-world deployment in unreliable distributed environments. 

\section{Programming Model}
\label{sec:programming_model}

Skywing's programming model separates algorithm-specific computation from the communication and execution mechanisms required to run decentralized algorithms. 
We refer to the Skywing runtime as the underlying software infrastructure that manages communication, message delivery, and distributed execution. 
The programming model provides the abstractions through which developers interact with this infrastructure while keeping algorithm-specific behavior encapsulated within processors.

%A primary objective of Skywing is to simplify the development of decentralized mathematical algorithms by separating algorithm logic from distributed systems concerns.
%Many distributed computing frameworks require developers to manage numerical updates, communication protocols, synchronization, and fault handling within the same implementation.
%This coupling increases implementation complexity and makes algorithms harder to reuse across deployment environments.
%Skywing addresses this challenge through a programming model built around three abstractions: \emph{Agent}, \emph{Processor}, and \emph{Iteration}. 
%We refer to the Skywing runtime as the underlying software infrastructure that manages communication, message delivery, and distributed execution on behalf of agents and processors. 
%The three programming abstractions separate communication, computation, and execution control while the runtime provides the infrastructure needed to support decentralized and asynchronous operation.
%The programming model is guided by three main goals: separating algorithm logic from communication infrastructure, supporting asynchronous execution in unreliable environments, and composing decentralized algorithms into larger workflows.
%Skywing provides a common execution model while allowing algorithm specific behavior to remain encapsulated within processors.

\subsection{Core Abstractions}
\label{sec:core_abstractions}
Skywing is organized around three core abstractions that define where distributed computation occurs, how local computation is expressed, and how execution progresses: \emph{Agent}, \emph{Processor}, and \emph{Iteration}. 
The \emph{Agent} represents a participant in a decentralized computation and serves as the local execution environment for one or more distributed algorithms. 
Applications configure an agent's neighbor relationships, while the runtime manages network connections and communication resources. 
Multiple algorithms may execute concurrently within an agent while maintaining independent computational state.
We use the term \emph{collective} to refer to a group of Skywing agents participating in a shared distributed computation. 
An individual agent maintains its local computational state and communicates with its configured neighbors, while the collective describes the set of agents over which the decentralized algorithm executes.

Internally, communication and algorithm execution operate in independent threads, allowing local computation to continue while messages are transmitted and received. 
The communication layer handles message transmission, reception, and delivery of published data, while processors access communicated values through Skywing's data interfaces rather than explicit socket operations.

The \emph{Processor} encapsulates the mathematical logic of a distributed algorithm. 
A processor maintains local state, consumes neighbor information through subscriptions, performs local computation, and prepares data for publication under named tags. 
Its update rule may depend on local state, recently available neighbor information, local data, and algorithm-specific parameters. 
This abstraction supports computational patterns ranging from consensus and optimization to iterative linear solvers, estimation, and control. 
In the Python interface, developers implement the algorithm-specific update and publication behavior while reusing the communication and execution infrastructure provided by Skywing.

The \emph{Iteration} combines an agent and a processor into an executing distributed task. 
During execution, it publishes processor data, collects newly available subscribed data, invokes the processor update rule, and evaluates stopping criteria. 
Iterations execute independently of the application thread, allowing applications to query intermediate results, update processor inputs, or launch additional computations while distributed algorithms continue running. 
The Python interface exposes these abstractions directly. 
At a high level, an application creates an agent, configures its neighbors, constructs an algorithm specific processor, and launches the processor through an iteration. 
For example, a Push Sum computation follows the pattern:

\begin{lstlisting}[style=skywing, caption={Basic Skywing execution pattern.},
	label={lst:basic_skywing}]
	agent = Agent(address, port)
	agent.configure_neighbors(neighbors)
	
	processor = PushSumProcessor(data=initial_value, N_k=N_k, agent=agent)
	
	iteration = Iteration(processor, agent)
	iteration.launch()
	
	estimate = iteration.query()
\end{lstlisting}

The application does not explicitly open sockets, send messages, or receive neighbor data. These operations are managed by the runtime, while the processor defines the algorithm specific update behavior and the iteration manages its execution. Applications can query the current result through the iteration interface and, when supported by the processor, provide updated local data through \texttt{update\_data()}. 

\begin{figure}[t]
	\centering
	\includegraphics[width=0.5\textwidth]{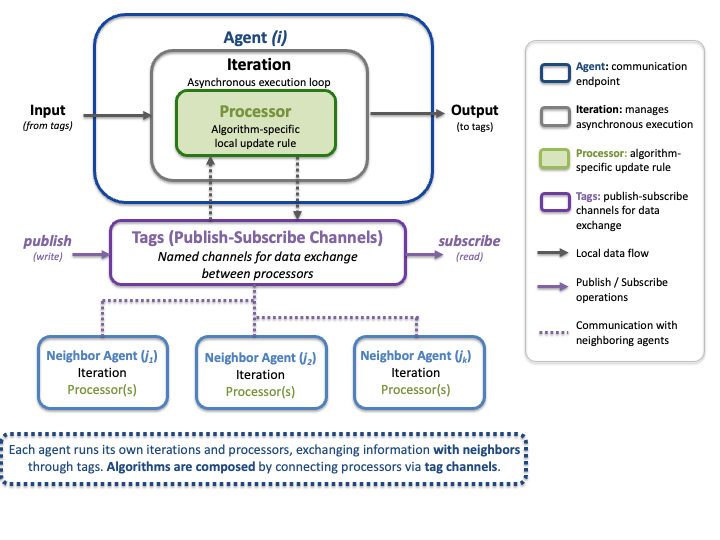}
\caption{Skywing programming model. An agent manages local communication and execution resources, an iteration manages asynchronous execution, and a processor implements the algorithm specific update rule. Processors exchange information through publish-subscribe tags while the runtime manages communication with neighboring agents.}
	\label{fig:programming_model}
\end{figure}

Figure~\ref{fig:programming_model} summarizes the relationship among these abstractions.
Each agent hosts one or more iterations, each iteration executes a processor, and processors exchange information through publish-subscribe tags.
The runtime manages communication and distributed execution, while processors define the algorithm-specific update rules.
This separation of concerns allows algorithm developers to focus on local mathematical computation while reusing the distributed communication and execution infrastructure provided by Skywing.

For multi-agent experiments and deployments, Skywing also provides driver utilities that automate common orchestration tasks. The driver infrastructure launches collections of agent processes, configures their communication topology and runtime parameters, and manages experiment outputs. These utilities sit above the Agent, Processor, and Iteration abstractions: they simplify launching a distributed computation but do not define the algorithm itself.

\subsection{Asynchronous Execution}

A defining feature of Skywing is its support for asynchronous execution.
Many distributed numerical algorithms are traditionally described using globally synchronized iterations, where every participant completes an update before any participant proceeds to the next step.
This model can be effective in controlled computing environments, but it becomes restrictive when communication delays, heterogeneous devices, and intermittent failures are expected.
In such settings, slow or temporarily unavailable participants can limit overall progress, and synchronization barriers can become a dominant source of overhead.

Skywing instead allows agents to execute independently and consume information as it becomes available.
Processors operate on the most recent neighbor data exposed through their subscriptions rather than assuming a globally consistent view of system state.
This execution model aligns with decentralized environments such as edge computing, autonomous systems, sensor networks, and distributed control, where global synchronization may be expensive, unreliable, or impossible.
By avoiding mandatory global synchronization, Skywing allows algorithms to continue making progress despite communication delays, variable compute rates, or temporary participant failures.

Skywing separates network level communication behavior from algorithm level synchronization policy.
Skywing implements network communication using nonblocking sockets so that communication with a slow or unreliable neighbor does not stall local execution. 
Communication and algorithm execution proceed independently, allowing an agent to continue local computation without waiting for every neighboring agent to communicate. 
As a result, agents may complete different numbers of local updates and may process different numbers of received messages over the same period of execution.
At the algorithm level, individual processor implementations can determine how much neighbor information is required before performing an update.
For example, a method may require particular neighbor updates before proceeding, while an asynchronous method may operate on the information currently available.
This distinction allows Skywing to support algorithms with different synchronization requirements without forcing blocking behavior into the communication layer.
The result is a flexible execution model in which the runtime provides nonblocking communication and data availability, while the algorithm determines how much coordination is needed before each update.

\subsection{Message Passing and Data Availability}

Skywing is designed to keep communicated information current while bounding communication and memory overhead. 
Processors exchange data through named publish subscribe tags. 
A tag identifies a logical stream of published values rather than an individual message: successive publications under the same tag represent updated values in that stream. 
Processors publish data under tags, and subscribing processors receive values associated with the tags to which they subscribe.

When values are published faster than they can be transmitted, Skywing retains the most recent unsent value for each tag. A newer publication therefore replaces an older unsent publication for the same tag. 
Communication buffers are also bounded; when communication buffers reach capacity, older queued data may be discarded in favor of newer data.
These policies prevent rapidly publishing agents from creating unbounded communication or memory overhead and favor current information over stale intermediate values.

Published data are serialized using MessagePack and represented using Pydantic based data models, providing consistent message formatting and validation across communication boundaries. 
These models support NumPy data, processor specific data, and application defined types, allowing processors to exchange structured data without implementing serialization logic themselves.

This behavior is particularly appropriate for iterative decentralized algorithms, where communicated values often represent evolving estimates, parameters, or partial solutions. 
In such settings, the most recent state is generally more useful than a backlog of intermediate states. 
Processors remain separate from these communication mechanisms: they define the data to publish and the computation to perform, while the runtime manages serialization, transmission, and communication buffering.

\subsection{Algorithm Composition}

\begin{figure*}[t]
	\centering
	\includegraphics[width=0.3\textwidth]{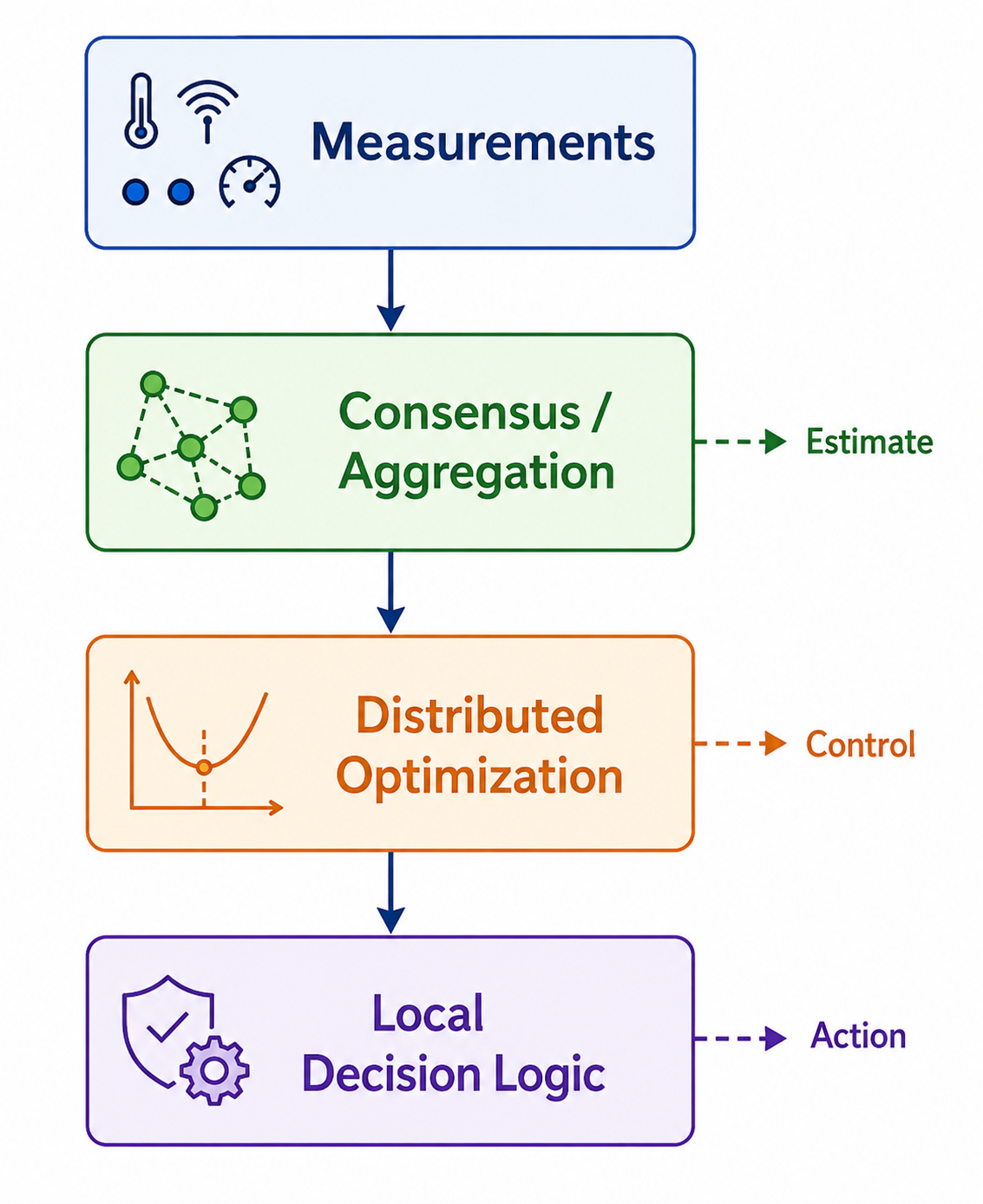}
	\caption{Example of algorithm composition in Skywing. Independent processors represent distinct computational stages that can be coordinated to form a larger decentralized workflow.}
	\label{fig:composition}
\end{figure*}

The separation between communication, computation, and execution also enables decentralized algorithms to be combined into larger workflows. 
Many real world applications require multiple computational stages rather than a single algorithm in isolation. 
For example, a distributed monitoring and control application may first aggregate measurements across a network, use the resulting estimate to compute a control signal, and then apply application specific decision logic before issuing an action.
In Skywing, each stage of such a workflow can be represented by an independent processor executing within an iteration. 
Processors encapsulate their own computational state and update rules, while the application coordinates the flow of data between stages. 
In the current Python interface, an application can query the output of one iteration and provide that value as input to another processor. 
This allows processors to remain separate computational components even when their outputs and inputs form a larger workflow.
Figure~\ref{fig:composition}  illustrates this composition model conceptually. 
A consensus or aggregation processor first combines local measurements to produce an estimate of a network quantity. 
A subsequent processor consumes that estimate and computes a control or optimization output. 
A final processor applies application specific decision logic before producing an action. 
The individual stages remain distinct processors, allowing their implementations to be developed and modified independently.

A simplified representation of the current composition pattern is:

\begin{lstlisting}[style=skywing   ,caption={Application-level composition of independent Skywing iterations.},
	label={lst:composition}]
	aggregation_processor = AggregationProcessor(measurement)
	control_processor = ControlProcessor(target)
	decision_processor = DecisionProcessor(parameters)
	
	aggregation_iter = Iteration(aggregation_processor, agent)
	control_iter = Iteration(control_processor, agent)
	decision_iter = Iteration(decision_processor, agent)
	
	aggregation_iter.launch()
	control_iter.launch()
	decision_iter.launch()
	
	while running:
		estimate = aggregation_iter.query()
		
		control_processor.update_data(estimate)
		control = control_iter.query()
		
		decision_processor.update_data(control)
		action = decision_iter.query()
\end{lstlisting}

The pseudocode highlights the separation between processor implementations and workflow orchestration. 
The aggregation processor does not need to implement the control or decision logic, and the downstream processors do not need to implement the aggregation algorithm. 
Instead, application level orchestration transfers data between independently executing processors. 
This structure allows individual stages to be replaced with alternative implementations provided that the expected data interface between stages is preserved.

Skywing's publish-subscribe communication model separately supports information exchange among distributed agents through tags. 
Together, processor based computation, iteration based execution, and tag based communication provide the building blocks for constructing larger decentralized applications. 
A natural extension of this model is to allow processors to consume the published outputs of other processors directly through tags, reducing the application level orchestration required to connect computational stages.
Section~\ref{sec:compose_workflow} demonstrates this composition pattern using a three stage monitoring and control workflow.

\subsection{Configuration and Design Implications}

The resulting programming model differs substantially from both traditional high-performance computing frameworks and workflow oriented distributed systems. 
Unlike MPI-style programming, Skywing does not expose low level communication operations as a primary programming construct. Developers express local update rules through processors, while communication and execution management are handled by the runtime. 
Likewise, Skywing differs from systems such as Hadoop and Spark, which are primarily designed for batch oriented data processing and centralized orchestration. 
Instead, Skywing focuses on continuously executing decentralized numerical methods whose behavior emerges from repeated local interactions among neighboring agents.

This separation of concerns isolates communication infrastructure from processor implementations. 
At the transport level, Skywing currently uses TCP based communication with nonblocking sockets, allowing local computation to continue without waiting for communication with slow or unreliable neighbors. 
The programming model is not inherently tied to TCP, however, and the separation between processor logic and communication infrastructure allows alternative transport mechanisms to be incorporated without changing algorithm specific processor implementations.
At the topology level, applications configure neighbor relationships. 
The runtime manages communication and execution, while processors define the algorithm specific update rules.
At the data level, processors publish and consume structured data associated with named tags. 
Message transmission and communication buffering are managed by the runtime rather than implemented separately by each processor.
This layered design allows processors with different mathematical objectives to reuse the same communication and execution infrastructure while keeping algorithm-specific logic separate from distributed systems concerns.

A consequence of this design is that new decentralized algorithms can often be incorporated by implementing their algorithm specific logic as processors while reusing Skywing's existing communication and execution infrastructure.
This reduces implementation effort, encourages algorithm reuse, and provides a common environment for evaluating different numerical methods. 
More broadly, the programming model shifts the focus of decentralized computing from repeatedly implementing communication infrastructure toward developing the mathematical algorithms themselves.
The separation between processor logic and runtime infrastructure also provides a foundation for future extensions. Additional configuration of publishing, buffering, and synchronization policies could be exposed without requiring those mechanisms to be implemented independently within each processor.
\subsection{Representative Algorithm Classes}
\label{sec:algorithms}

The Skywing programming model is intended to support a range of decentralized mathematical methods rather than a single algorithm family. 
Current implementations span consensus, optimization, and numerical linear algebra, as summarized in Table~\ref{tab:representative_algorithms}. 
These categories differ in their mathematical objectives, state representations, and communication patterns, but they are expressed using the same Agent, Processor, and Iteration abstractions.

\begin{table}[t]
	\centering
	\caption{Representative algorithms implemented in Skywing.}
	\label{tab:representative_algorithms}
	\begin{tabular}{lll}
		\hline
		Algorithm & Category & Data Exchanged \\
		\hline
		Max Consensus & Consensus & Neighbor estimates \\
		Push Sum & Consensus & Estimates and weights \\
		SGD & Optimization & Model parameters \\
		ADMM & Optimization & Primal and dual variables \\
		Jacobi & Linear solver & Partial solutions \\
		s-ACD & Linear solver & Intermediate iterates \\
		\hline
	\end{tabular}
\end{table}

Consensus and aggregation methods provide the simplest examples of the programming model.
Algorithms such as Max Consensus and Push Sum maintain local estimates that are repeatedly exchanged among neighboring agents until a global quantity is recovered or approximated.
These methods demonstrate decentralized information exchange without centralized coordination and provide useful building blocks for distributed estimation, resource allocation, and control.

Distributed optimization methods introduce richer local state and more structured update rules.
Implementations such as SGD and ADMM require agents to maintain optimization variables, exchange parameter information, and update local state based on both local objectives and neighbor data.
Their inclusion demonstrates that the same execution model used for consensus can also support algorithms with local objectives, constraints, and coordinated updates.

Skywing also includes representative decentralized numerical linear algebra processors, including Jacobi iterative methods and s-step asynchronous conjugate direction methods.
These methods operate on structured matrix and vector data and expose challenges such as stale neighbor information, asynchronous progress, and corrupted intermediate values.
They therefore provide useful examples of how the Skywing programming model can support scientific computing style algorithms in decentralized and unreliable environments.

Together, these implementations show that the Agent, Processor, and Iteration abstractions accommodate substantially different computational patterns, state representations, and communication requirements across consensus, optimization, and numerical linear algebra.

\section{Case Studies and Numerical Demonstrations}
\label{sec:demonstration}

The objective of this evaluation is to demonstrate the range of decentralized mathematical computations that can be implemented and studied within the Skywing programming model. 
The experiments are intended as demonstrations of the platform rather than benchmarks of individual algorithms. 
They exercise different aspects of the programming model, including decentralized execution, processor composition, and algorithmic resilience under unreliable conditions.

\subsection{Experimental Setup}

All experiments in this section execute using the native Skywing runtime.
Each agent executes as an independent operating system process and communicates with neighboring agents through Skywing's TCP based networking layer.
Consequently, the reported results include the effects of interprocess communication, asynchronous execution, operating system scheduling, and network latency.
The experiments therefore evaluate the complete software platform rather than isolated algorithm implementations.
Unless otherwise stated, experiments use either a ring or line communication topology.
Each experiment is repeated over multiple runtime trials.
Unless otherwise stated, reported trajectories correspond to mean behavior across trials; exceptions are identified in the individual experiments.
The experiments are orchestrated through Skywing's driver infrastructure described in Section~\ref{sec:core_abstractions}, which launches
agent collectives and configures communication topologies and runtime parameters. 
Table~\ref{tab:experiment_summary} summarizes the demonstrations used in this section.

\begin{table}[t]
	\centering
	\caption{Summary of representative Skywing demonstrations.}
	\label{tab:experiment_summary}
	\begin{tabular}{p{0.42\linewidth}p{0.14\linewidth}p{0.18\linewidth}p{0.14\linewidth}}
		\hline
		Experiment & Agents & Topology & Trials \\
		\hline
		Push Sum consensus & 8 & Ring & 5 \\
		Max consensus & 8 & Line & 5 \\
		Composed workflow & 8 & Ring & 5 \\
		Robust Push Sum & 8 & Ring & 5 \\
		ASJ / ASJ-R & 16 & Line & 5 \\
		\hline
	\end{tabular}
\end{table}
%
%The selected experiments are intended to exercise different capabilities of the programming model rather than to provide an exhaustive benchmark suite.
%The consensus experiments establish that different decentralized algorithms can be expressed using the same abstractions.
%The composed workflow demonstrates that processors can be assembled into larger applications without modifying their implementations.
%The resilience experiments demonstrate that delayed communication and data corruption can be studied within the same runtime while leaving the communication infrastructure unchanged.

\subsection{Consensus}
We begin with two consensus algorithms that exercise the same Skywing programming abstractions while exhibiting different information propagation behavior. 
Push Sum estimates a network average through repeated exchange of weighted state variables, while Max Consensus propagates the largest observed value through the communication graph. 
Together, these examples demonstrate that the Agent, Processor, and Iteration abstractions are not tied to a particular consensus update rule.

For Push Sum, the trial-level consensus error is measured as the maximum absolute deviation from the true network average, 
\[
e(t) = \max_i \left|x_i(t) - x^*\right|,
\]
where $x_i(t)$ is the estimate maintained by agent $i$ at time $t$ and $x^*$ is the true network average. The error is computed independently for each trial and then averaged across the five trials.

\begin{figure}[h!]
	\centering
	\includegraphics[width=.7\linewidth]{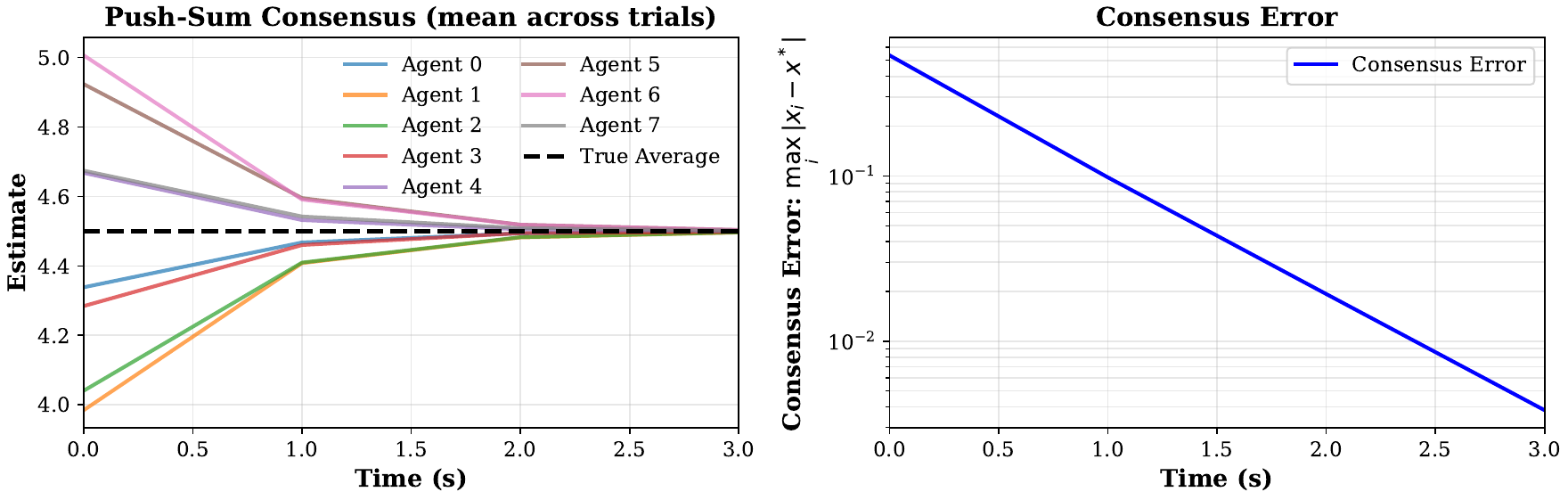}
   \caption{Push Sum averaging executed using the native Skywing runtime. The left panel shows mean agent trajectories across five trials, while the right panel shows the mean trial level consensus error.}
	\label{fig:pushsum_runtime}
\end{figure}

The Skywing Push Sum implementation follows the algorithmic formulation described in~\cite{tian_push_sum}, with the local update rule encapsulated by a Push Sum processor. 
For this experiment, eight agents are connected in a bidirectional ring topology and initialized with distinct scalar values. Figure~\ref{fig:pushsum_runtime} shows the resulting convergence. 
The agent estimates converge toward the network average, while the consensus error decreases toward zero.

Max Consensus provides a complementary example with a different information propagation pattern. 
To make the propagation dynamics observable over the runtime of the experiment, we use a gossip-based Max Consensus processor that maintains separate communicated and reported values. 
Each agent communicates the largest value it has observed, so the communicated state follows the standard max update even though the reported estimate is damped for visualization.
For reporting, the processor maintains a damped estimate that approaches the largest value currently known by the agent according to
\[
d_i^{(k+1)}
=
(1-\alpha)d_i^{(k)}
+
\alpha m_i^{(k)},
\]
where $m_i^{(k)}$ is the largest value known by agent $i$, $d_i^{(k)}$ is its reported estimate, and $\alpha=0.15$ is the damping factor. The damping affects only the reported value and does not alter the values communicated among
agents.

\begin{figure}[h!]
	\centering
	\includegraphics[width=.7\linewidth]{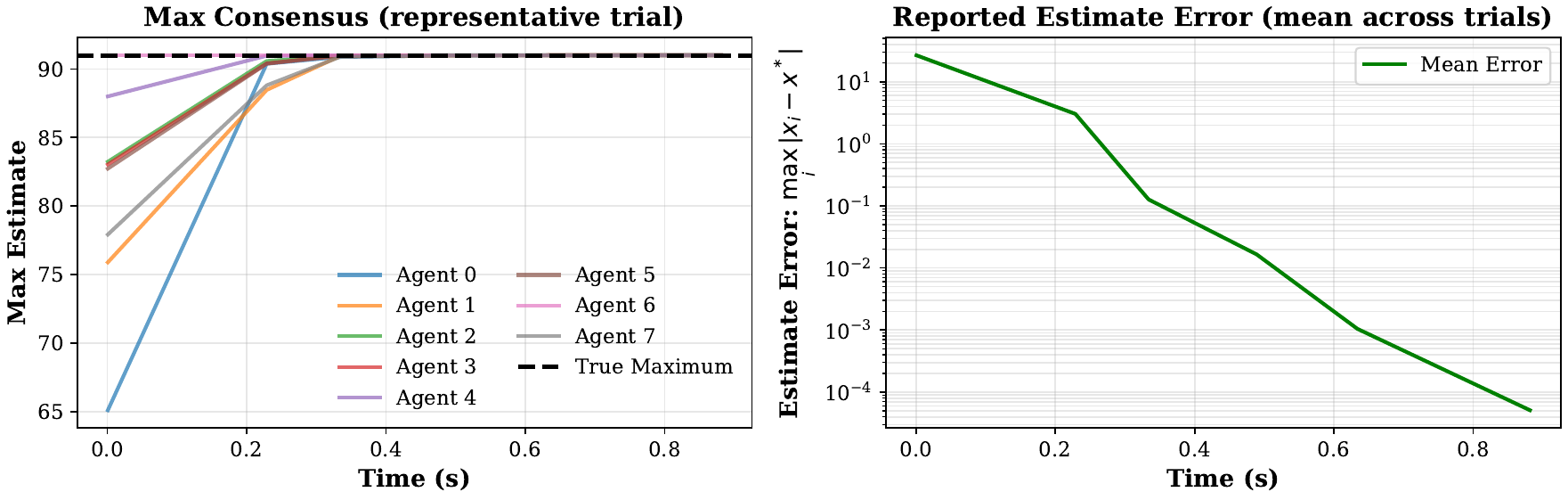}
	\caption{Gossip-based Max Consensus executed within Skywing. The left panel shows the reported estimates for a representative trial, while the right panel shows the mean reported-estimate error across five trials.
		Agents communicate their current known maximum directly, while a damping factor is applied only to the reported estimates to make the convergence
		dynamics visible.}
	\label{fig:max_consensus_runtime}
\end{figure}

Eight agents are arranged in a bidirectional line topology so that information about the maximum must propagate through successive neighbors.
Figure~\ref{fig:max_consensus_runtime} shows one representative execution in the left panel and the mean reported-estimate error across five trials in the right panel. 
The gradual changes in the plotted trajectories reflect the damping applied to the reported estimates, while the underlying maximum is communicated directly among neighboring agents. For each trial, the reported-estimate error is computed as
\[
e_r(t)=\max_i \left|d_{i,r}(t)-x_{\max}\right|,
\]
where $d_{i,r}(t)$ is the reported estimate of agent $i$ at time $t$ in trial $r$, $x_{\max}$ is the true maximum of the initial agent values, and $e_r(t)$ is the maximum reported-estimate error across agents for that trial. 
The right panel reports the mean of $e_r(t)$ across the five trials.

\subsection{Composable Distributed Workflows}
\label{sec:compose_workflow}
Many decentralized applications require multiple computations to operate together rather than a single algorithm in isolation. 
For example, a distributed energy management application may aggregate measurements, compute control actions, and apply safety constraints before issuing local commands. 
To demonstrate how independent Skywing processors can be combined into a larger application, we construct a three-stage decentralized monitoring and control workflow inspired by distributed energy resource coordination.

Each agent executes three independent processors through separate Skywing iterations. The first stage estimates the average network measurement using Push Sum. 
The second stage computes a proportional control signal from the aggregated estimate. The third stage applies a safety margin and limits the change in the control action before producing the final decision. 
Application-level orchestration connects the stages by querying the output of one iteration and providing that value as input to the next processor.

The workflow is summarized by
\[
\begin{aligned}
	\hat{x} &= \operatorname{PushSum}(m_i),
	&& \text{Aggregation}\\
	u &= \operatorname{clip}\left(K(r-\hat{x}),-20,20\right),
	&& \text{Control}\\
	a_k &= a_{k-1}
	+\operatorname{clip}\left(0.9u-a_{k-1},-3,3\right),
	&& \text{Decision}
\end{aligned}
\]
where $m_i$ denotes the local measurement, $\hat{x}$ denotes the distributed estimate of the network average, $r$ denotes the target value, $K=1$ denotes the proportional gain, and $a_k$ denotes the final decision variable.

To test dynamic adaptation, the local measurements are changed during execution. 
Initially, each agent observes a fixed value sampled between 90 and 100. 
At iteration 10, these measurements are replaced by fixed values sampled between 110 and 115 while the workflow continues running. 
The updated measurements first affect the distributed aggregation stage. 
The resulting estimate is then passed to the control processor, and the control signal is subsequently passed to the decision processor.

\begin{figure}[h!]
	\centering
	\includegraphics[width=.7\linewidth]{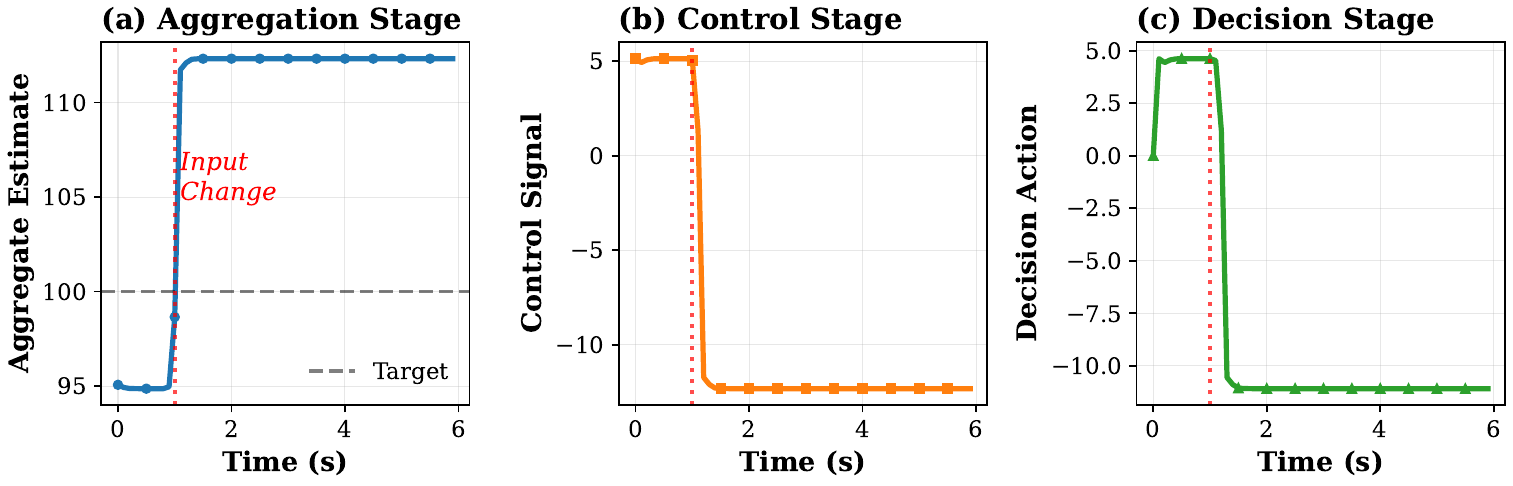}
	\caption{Three-stage composed workflow implemented within Skywing. A distributed aggregation stage feeds a local control stage and a decision stage, and a change in the input measurements propagates through the workflow without modifying processor implementations.}
	\label{fig:composed_workflow_runtime}
\end{figure}

Figure~\ref{fig:composed_workflow_runtime} shows the resulting behavior.
The aggregation stage moves toward the updated network average after the input change.
The control stage responds by generating a negative control signal as the estimate rises above the target.
The decision stage applies the safety margin and limits how much the action can change between successive updates.
In a power systems setting, this final stage can represent local operating constraints that limit abrupt changes in local actuation.
The example therefore captures a simplified monitoring and control pipeline in which network-wide information is first estimated, translated into a control response, and then adjusted according to local operating constraints.

Although simplified, the workflow demonstrates how distinct computational stages can be combined within Skywing while remaining separate processor implementations.

The aggregation stage could be replaced by another consensus or estimation processor without changing the control or decision logic, provided that the expected data interface is preserved.
Likewise, the downstream stages could be replaced independently.
This modular structure allows individual processors to remain focused on their own computational role while participating in a larger composed workflow.

\subsection{Resilience Under Unreliable Communication}
A central motivation for Skywing is the need to study decentralized algorithms in environments where communication may be delayed or unreliable.
Skywing provides asynchronous communication and execution support, but resilience to delayed or corrupted information is primarily an algorithmic property. 
The platform therefore provides a common environment in which standard and resilient processors can be compared under identical runtime conditions.

We compare standard Push Sum with a resilient Push Sum implementation based on a counter mechanism for delayed messages~\cite{olshevsky2018fully_async_pushsum}. 
For both algorithms, eight agents are arranged in a bidirectional ring topology and initialized with identical local values. Each experiment is repeated over five runtime trials. 

A deterministic receive side delay schedule is generated and applied across the experiments. 
For each sender receiver pair and sampled iteration, received neighbor data are withheld with probability 0.70 for a randomly selected duration of up to thirty iterations. 
The delay model is implemented at the processor level using a reusable delay wrapper.
After Skywing receives neighbor data, the wrapper either makes the data immediately available to the underlying processor or holds it in a local buffer until the prescribed release iteration.
The standard and resilient Push Sum processors therefore retain their original update rules, and the experiment does not modify Skywing's underlying TCP communication layer. 
The same delay schedule is applied to both processor implementations, providing a paired comparison under identical delay conditions.

\begin{figure}[h!]
	\centering
	\includegraphics[width=.5\linewidth]{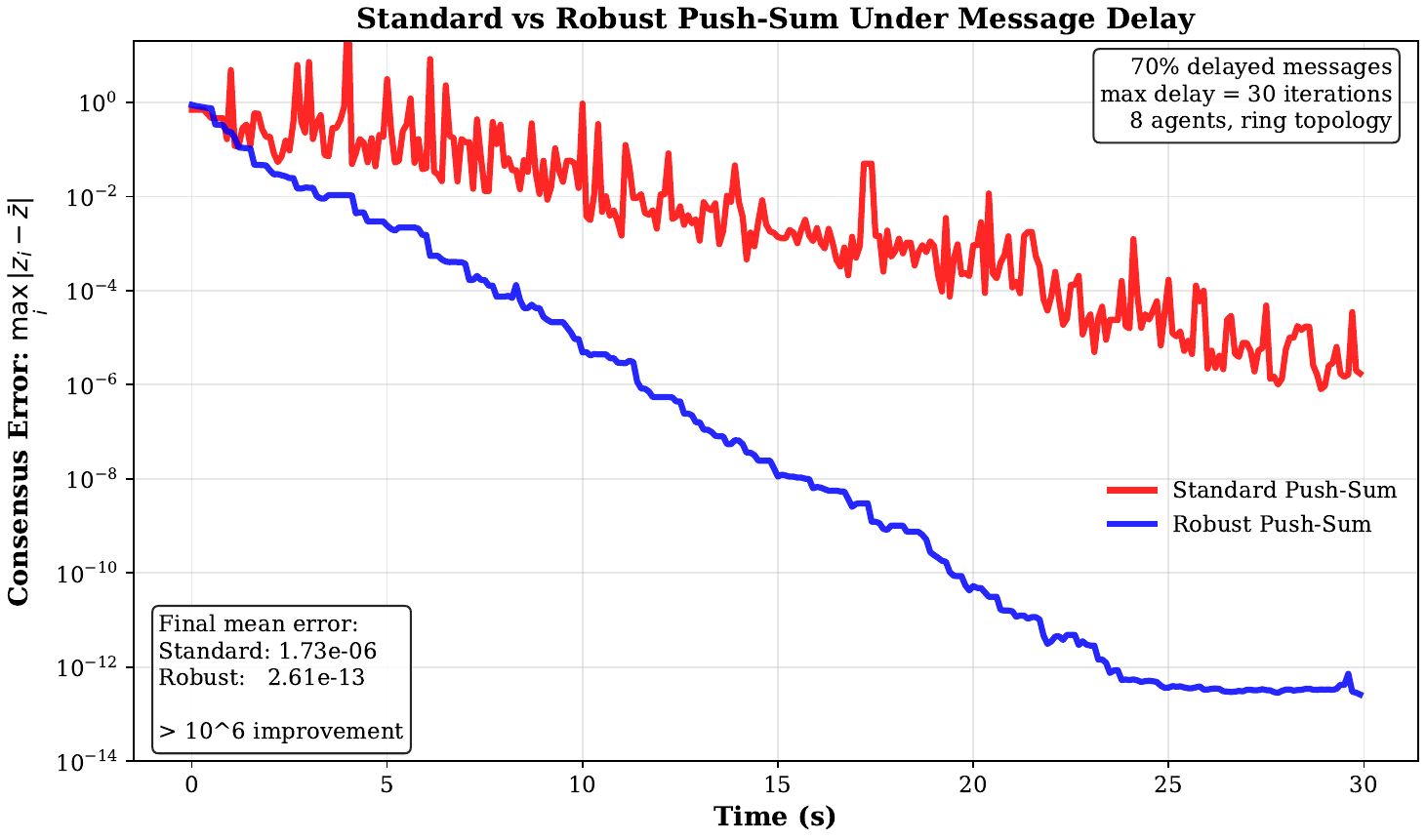}
	\caption{Comparison of standard and resilient Push Sum under identical delayed communication conditions. Both algorithms execute within the same Skywing runtime, so differences in convergence arise from the processor implementations.}
	\label{fig:robust_pushsum_runtime}
\end{figure}

Figure~\ref{fig:robust_pushsum_runtime} compares the resulting convergence behavior. 
Both methods initially reduce the consensus error. 
As delayed information accumulates, the standard Push Sum implementation degrades, while the resilient implementation continues reducing the error. 
This experiment illustrates how Skywing can be used to evaluate algorithmic resilience under controlled communication delays while reusing the same communication and execution infrastructure.

\subsection{Resilient Numerical Linear Algebra}
We next consider decentralized iterative linear solvers, which involve richer local state and more structured communication than consensus algorithms.
These methods maintain distributed approximations to the solution of a global linear system while exchanging partial solution information among neighboring agents. 
They therefore provide a more demanding test of Skywing's ability to support scientific computing style algorithms in decentralized and unreliable environments.

We compare standard asynchronous Jacobi \cite{wolfsonpou2018convergence}, denoted ASJ, with a resilient variant \cite{vogl2024async_jacobi_resilience}, denoted ASJ-R, under a controlled malevolent corruption model.
Both algorithms execute using Skywing's native linear solver driver. 
The corruption model is implemented at the processor level through a reusable mixin that modifies data during publication. 
The underlying ASJ and ASJ-R update rules remain unchanged; when the selected agent is in a degraded
state, the mixin perturbs the solution values prepared for publication before they are communicated to neighboring agents. 
The agent's locally stored solution state is not modified. 
This design allows corruption models to be introduced or replaced through processor-level publication behavior without changing Skywing's communication infrastructure.

Each experiment uses sixteen agents collaboratively solving a distributed graph Laplacian system with 400 unknowns. 
Communication occurs over a line topology, and each experiment is repeated over five trials. 
Agent 9 is selected as the corrupted agent. 
Corruption occurs periodically: the agent is degraded for 5 iterations during every 50-iteration period. 
While degraded, the values it publishes are perturbed by offsets sampled independently from a normal
distribution with mean $\delta=0.2$ and standard deviation $0.5|\delta|$.

A different random seed is used for each trial to generate the corruption offsets. Within each trial, the same corruption parameters and random seed are used for ASJ and ASJ-R, providing a paired comparison under equivalent corruption conditions. 
This design isolates the effect of the resilience mechanism from differences in runtime infrastructure or experimental conditions.

\begin{figure}[t]
	\centering
	\includegraphics[width=.5\linewidth]{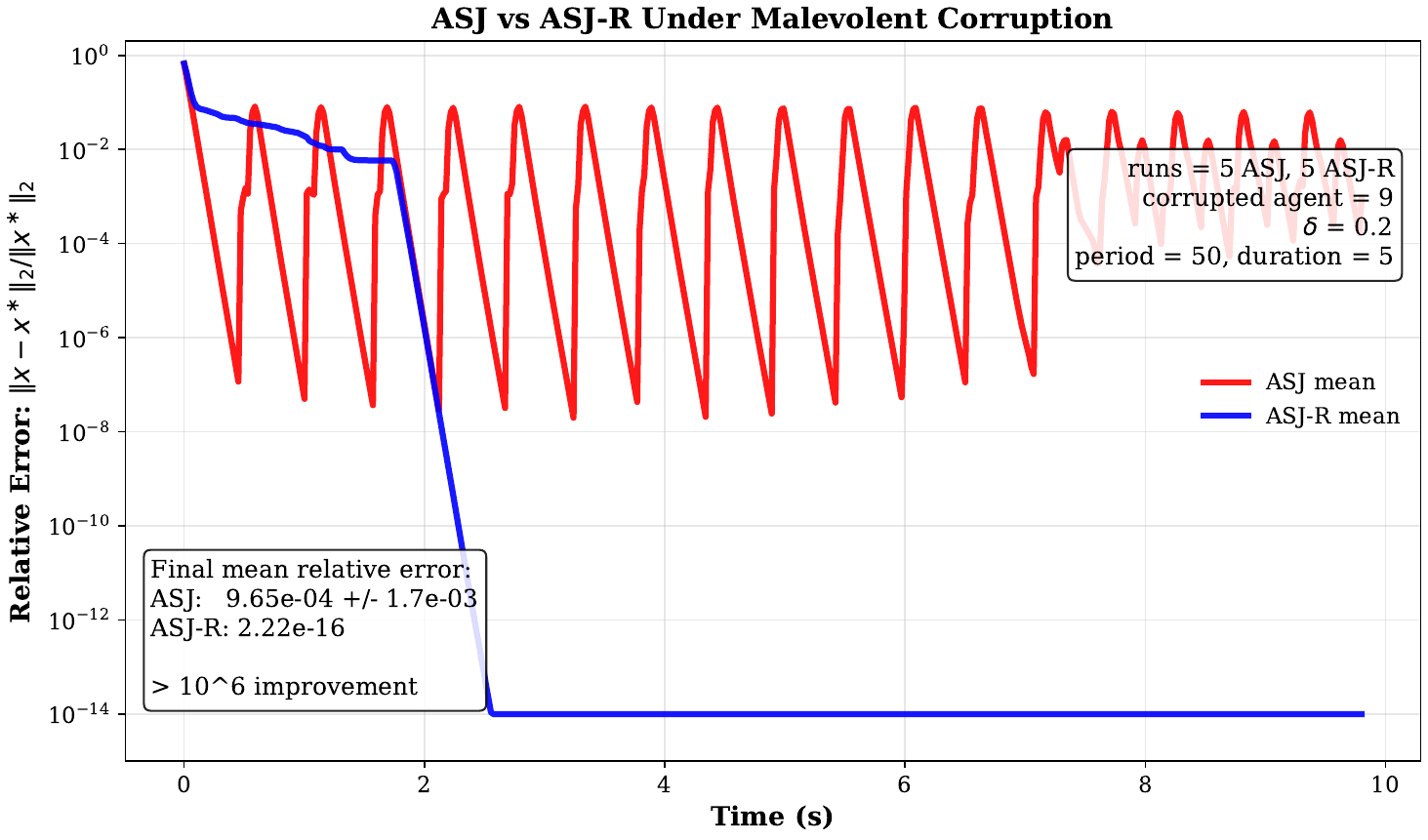}
   \caption{Comparison of asynchronous Jacobi and resilient asynchronous Jacobi under periodic malevolent corruption. Both algorithms execute within the same Skywing runtime and experience the same corruption schedule. }
	\label{fig:jacobi_malevolent_comparison}
\end{figure}

Figure~\ref{fig:jacobi_malevolent_comparison} summarizes the resulting behavior.
Both methods initially reduce the relative solution error. 
As corruption accumulates, the standard ASJ implementation loses accuracy, while ASJ-R  continues reducing the error toward machine precision. 
The experiment demonstrates that Skywing can support algorithm-specific resilience mechanisms while preserving the same communication and execution infrastructure used by the other demonstrations.

\subsection{Discussion}

The demonstrations in this section illustrate Skywing's flexibility across several dimensions of decentralized mathematical computing.
The consensus experiments show that different decentralized algorithms can be implemented using the same programming abstractions and runtime infrastructure.
The composed workflow shows that independent processors can be coordinated within Skywing to form a larger computational workflow while remaining separate modular components.
The resilience experiments show that delayed communication and adversarial data corruption can be studied within the same execution environment while leaving the runtime infrastructure unchanged.
These results should be interpreted as platform demonstrations rather than competitive algorithm benchmarks.
The goal is not to show that a particular implementation of Push Sum, Max Consensus, or asynchronous Jacobi outperforms all alternatives.
The goal is to show that diverse decentralized mathematical methods can be implemented, executed, composed, and evaluated within a common software environment.
In this sense, Skywing serves both as a deployment framework for decentralized applications and as a research platform for developing resilient mathematical algorithms in unreliable computing environments.

\section{Conclusion}

This paper presented Skywing, an open source platform for decentralized mathematical computing in unreliable environments.
Skywing was developed in response to a growing gap between traditional distributed computing frameworks and the requirements of emerging decentralized systems.
While high-performance computing environments often emphasize synchronization, reliable communication, and centralized resource management, many modern applications operate under substantially different assumptions.
Edge computing platforms, autonomous systems, distributed sensing applications, and cyber physical infrastructures frequently involve heterogeneous devices connected through unreliable communication networks.
In such environments, strict synchronization may be impractical, and centralized coordination may be undesirable or impossible.

The design of Skywing reflects these requirements.
Rather than optimizing for tightly coupled numerical computation on reliable infrastructure, Skywing prioritizes decentralization, asynchronous execution, and algorithm composability.
The platform separates algorithm logic from distributed systems concerns through the \emph{Agent}, \emph{Processor}, and \emph{Iteration} abstractions.
Agents provide the local execution environment, processors encapsulate algorithm specific update rules, and iterations manage the execution of distributed tasks.
Together, these abstractions allow decentralized algorithms to be implemented as reusable computational components while the runtime manages communication and distributed execution.

A central contribution of Skywing is the programming model itself.
Existing frameworks often require algorithm developers to intertwine mathematical logic with communication and execution concerns.
In contrast, Skywing allows researchers to focus on algorithm design while relying on a common execution substrate for communication and coordination.
This separation improves modularity, portability, and code reuse, while also enabling decentralized algorithms to be evaluated and compared within a shared framework.
The inclusion of consensus methods, optimization algorithms, composed workflows, and iterative linear solvers demonstrates that the abstractions support a broad range of computational patterns.

The numerical demonstrations show that the same Skywing runtime can support several classes of decentralized mathematical methods.
Push Sum and Max Consensus illustrate decentralized information exchange.
The composed monitoring and control workflow shows how independent processors can be coordinated within Skywing to form a larger computational workflow while remaining separate modular components.
The resilient Push Sum and asynchronous Jacobi experiments demonstrate that delayed communication and data corruption can be studied within the same execution environment.
These examples illustrate that Skywing is not tied to a single algorithm or application domain, but instead provides reusable infrastructure for implementing, composing, and evaluating decentralized methods under unreliable operating conditions.

Skywing also involves deliberate tradeoffs.
By emphasizing abstraction and composability, the platform may not match the communication efficiency of specialized implementations designed for a single algorithm or deployment environment.
The publish-subscribe model introduces additional indirection compared with explicit message passing.
The Python first design lowers the barrier to algorithm development and experimentation, but may limit performance for some workloads.
These tradeoffs are intentional: Skywing is not intended to replace highly optimized HPC software stacks, but to simplify the development and deployment of decentralized algorithms in environments where reliability and synchronization cannot be assumed.

Several limitations remain in the current implementation.
Neighbor relationships are currently configured explicitly at deployment time, which limits support for highly dynamic networks.
The communication layer assumes cooperative participants and does not by itself provide Byzantine fault tolerance or protection against malicious agents.
The current platform also focuses on decentralized execution rather than persistent storage, long term state management, or large scale workflow orchestration.
These limitations reflect the present scope of the project rather than fundamental constraints of the programming model.

Future work will extend Skywing in several directions.
Planned developments include adaptive and self organizing communication topologies, stronger resilience mechanisms for adversarial behavior and corrupted data, additional communication backends, richer buffering and scheduling policies, and broader libraries of optimization, estimation, and numerical linear algebra processors.
We also plan to apply Skywing to larger application driven workflows in power systems, autonomous sensing, and edge computing.
More broadly, we view Skywing as a step toward providing decentralized mathematical computing with reusable software abstractions and infrastructure comparable to those available in traditional scientific computing. 
Just as traditional scientific computing has benefited from common abstractions and reusable numerical libraries, decentralized computing may benefit from platforms that allow algorithms to be developed, composed, and deployed independently of the details of the underlying communication infrastructure.

\begin{acks}
	This work was supported by the LLNL-LDRD Program under Project No.\ 21-FS-007,  22-ERD-045, and 24-ERD-030. 	This work was performed under the auspices of the U.S. Department of Energy by Lawrence Livermore National Laboratory under Contract DE-AC52-07NA27344. 	LLNL-JRNL-2023583.
\end{acks}

\bibliographystyle{ACM-Reference-Format}
\bibliography{acmart}

\end{document}